\documentstyle[psfig,nato,numreferences]{crckapb}
\begin{opening}
\title{Macroscopic quantum tunneling in magnetic nanostructures}
\author{Boris A. Ivanov}
\author{Alexei K. Kolezhuk}
\institute{Institute of Magnetism,
 National Academy of Sciences \\
and Ministry of Education of Ukraine\\
 36(B) Vernadskii avenue, Kiev 252142, Ukraine}
\end{opening}

\begin{document}
\maketitle
\begin{abstract}
Theoretical foundations of the problem of quantum spin tunneling in magnetic
nanostructures are presented. Several model problems are considered in detail,
including recent new results on tunneling in antiferromagnetic nanoparticles
and  topologically nontrivial magnetic structures in systems with reduced
dimension. 
\end{abstract}

\section{Introduction}

It is well known that magnetic ordering is an essentially quantum
phenomenon. According to the Bohr -- van Leeven theorem (see, e.g.,
\cite{Vonsovskii}), the magnetization of a thermodynamically equilibrium
classical system of charged particles is zero even in presence of an external
magnetic field. Classical theories of magnetic properties were based on
certain assumptions going beyond the limits of classical physics (e.g., the
existence of stable micro-particles with nonzero magnetic moment assumed in
Langevin's theory of paramagnetism \cite{Vonsovskii}). The nature of magnetic
ordering was revealed only after the discovery of modern quantum mechanics in
the works of Heisenberg, Frenkel and Dorfman. In 30s, many remarkable results
were obtained within the microscopic quantum theory: Bloch \cite{Bloch30}
predicted the existence of magnons and low-temperature behavior of
magnetization; Bethe \cite{Bethe31} was able to construct the complete set of
excited states for a spin-$1\over2$ chain, including nonlinear soliton-type
excitations (spin complexes). 

The `undivided rule' of the quantum
theory of magnetism lasted only till 1935, when in the well-known work Landau
and Lifshitz \cite{LL} formulated the equation describing the dynamics of
macroscopic magnetization of a ferromagnet (FM). When deriving the
Landau-Lifshitz (LL) equation, a quantum picture of magnetic ordering was
used, particularly, the exchange nature of spin interaction, but the LL
equation itself has the form of a classical equation for the magnetization
$\vec{M}$. Later on the basis of the LL equation the macroscopic theory of
magnetism was developed and enormous number of various phenomena were
described \cite{SW,BICG} (an overview of modern phenomenological theory of
magnetically ordered media can be also found in this book in the lecture by
V.G.Bar'yakhtar). 

This lecture presents an introduction to the foundations of a new,
fast-developing topic in the physics of magnetism, Macroscopic Quantum
Tunnelling (MQT). Let us first address briefly the scope of problems belonging
to this field. MQT problems can be roughly divided into two main types. First
of all, there are phenomena connected with the underbarrier transition from a
metastable state, corresponding to a local minimum of the magnet energy, to a
stable one. Such effects were observed in low-temperature remagnetization
processes in small FM particles as well as in macroscopic samples (due to the
tunneling depinning of domain walls), see the recent review
\cite{TejadaZhang95}. Such phenomena of ``quantum escape'' are typical not
only for magnets, e.g., quantum depinning of vortices contributes
significantly to the energy losses in HTSC materials \cite{HTSC}. 

Here we will concentrate on another type of phenomena, the so-called {\em
coherent MQT.\/} To illustrate their main feature, let us consider a small FM
particle with the easy axis along the $Oz$ direction. If the particle size is
small enough (much less than the domain wall thickness $\Delta_{0}$), the
particle is in a single-domain state, because the exchange interaction makes
the appearance of a state with magnetic inhomogeneities energetically
unfavorable. Then, from the point of view of classical physics, the ground
state of the particle is twofold degenerate. Those two states correspond to
two local minima of the anisotropy energy and are macroscopically different
since they have different values of macroscopic magnetization $\vec{M}=\pm
M_{0} \vec{e}_{z}$. The situation is the same as in the elementary mechanical
problem of a particle in two-well potential $U(x)$ having equivalent minima at
$x=\pm a$, see Fig.\ \ref{fig:two-well}. In classical mechanics the minimum of
energy corresponds to a particle located in one of the two local minima of the
potential. 

However, from quantum mechanics textbooks it is well known that the actual
situation is {\em qualitatively\/} different: the particle is ``spread'' over
two wells, and the ground state is nondegenerate \cite{LL-QM}. One can expect
that the same should be true for a FM particle: its correct ground state will
be a superposition of ``up'' and ``down'' states, and the mean value of
magnetization will be zero. Such picture was first proposed by Chudnovsky
\cite{Chud79}; further calculations showed \cite{ChudGunther88} that such
effects are possible for FM particles with rather large number of spins (about
$10^{3}\div10^{4}$). The tunneling effects, according to the theoretical
estimates \cite{BarbaraChud90,KriveZaslavskii90}, should be even more
important for small particles of antiferromagnet (AFM); the effects of quantum
coherence in AFM particles were observed in Ref.\ \cite{Awschalom+92}.

Thus, an important feature of quantum mechanics, a possibility of underbarrier
transitions, can manifest itself in magnetic particles on a macroscopic
(strictly speaking, mesoscopic) scale. Maybe even more interesting is the
manifestation of another characteristic feature of quantum physics, viz.\ the
effects of quantum interference. Such effects arise in the problem of MQT in
magnetic nanostructures and can partially or completely suppress tunneling,
restoring the initial degeneracy of the ground state
\cite{Loss+92,DelftHenley92}. We wish to remark that understanding that
motion of particles along very different classical trajectories can ``sum up''
in some sense and yield an interference picture was one of the crucial points
in the development of quantum mechanics, and a considerable part of the
well-known discussion between Bohr and Einstein was devoted to this problem.
Besides the importance of the tunneling phenomena in magnets from the
fundamental point of view, they are potentially important for the future
magnetic devices working on a nanoscale. 

In the present lecture we restrict ourselves to discussing the problems of
coherent MQT in various mesoscopic magnetic structures. The paper is organized
as follows: Sect.\ \ref{sec:basics} contains the elementary description of the
instanton formalism, traditionally used in the theoretical treatment of MQT
problems. Since the instanton approach, though being the most straightforward
one, is based on rather complicated mathematical formalism, we will discuss it
in parallel with simple and widely known semiclassical approximation of
quantum mechanics. The point is that those two approaches are equally adequate
for treating the problem of MQT in small particles, and the ``standard''
semiclassical calculations, easily reproducible by anybody who learned
foundations of quantum mechanics, may be helpful for understanding the
structure of the results derived within the instanton technique.  Further, in
Sections \ref{sec:FM} and \ref{sec:AFM} we discuss the problem of MQT in
ferro- and antiferromagnetic small particles, with a special attention to the
interference effects.  For the description of AFM we use simple but adequate
approach based on the equations for the dynamics of the antiferromagnetism
vector $\vec{l}$. This approach easily allows one to keep trace of the actual
magnetic symmetry of the crystal; the symmetry is lowered when external
magnetic field is applied or when certain weak interactions, e.g., the
so-called Dzyaloshinskii-Moriya (DM) interaction, are taken into account,
which leads to quite nontrivial interference phenomena. Section \ref{sec:top}
is devoted to the analysis of coherent MQT in ``topological nanostructures,''
i.e.\ static inhomogeneous states of magnets with topologically nontrivial
distribution of magnetization; among the examples considered there are domain
walls in one-dimensional (1D) magnets \cite{ivkol94,ivkol95tun,ivkol96jetp},
magnetic vortices \cite{GalkinaIv95} and disclinations \cite{IvKir97} in 2D
antiferromagnets, and antiferromagnetic rings with odd number of spins
\cite{kireev}. For those problems, when the description of tunneling involves
multidimensional (space-time) instantons, there is no alternative to the
instanton approach and its use is decisive.
Finally, Section \ref{sec:summary} contains a brief summary and discussion of
several problems which are either left out of our consideration or unsolved.

\section{Basics of Tunneling: With and Without Instantons}
\label{sec:basics}

For the sake of the presentation completeness, let us recall briefly the main
concepts of the instanton technique, since we will extensively use them below.

In quantum field theory, the propagator, i.e., the amplitude of probability
$P_{12}$ of the transition from any given state with the field configuration
$\varphi_{A}(x)$ at $t=0$ to another state $\varphi_{B}(x)$ at $t=t_{0}$ is
determined by the path integral
\begin{equation}
  \label{1to2}
  P_{AB}=\langle \varphi_{A}|e^{i \widehat{H}t_{0}/\hbar}|\varphi_{B}\rangle
=\int_{\varphi(x,0)=\varphi_{A}(x)}^{\varphi(x,t)=\varphi_{B}(x)}
 {\cal D}\varphi(x,t )\, \exp\big\{i{\cal A}[\varphi]/\hbar \big\}\,,
\end{equation}
where 
\[
{\cal A}[\varphi]=\int_{0}^{t_{0}} dt\int dx\, {\cal L}[\varphi(x,t)] 
\]
is the action functional. Here ${\cal L}$ is the Lagrangian density, and the
integration in (\ref{1to2}) goes over all space-time field configurations
$\varphi(x,t)$ satisfying the boundary conditions
$\varphi(x,0)=\varphi_{A}(x)$ and $\varphi(x,t_{0})=\varphi_{B}(x)$.  (We leave
out the problem of a consistent definition of the measure ${\cal D} \varphi$
that arises for systems with infinitely many degrees of freedom, keeping in
mind that we are going to talk about the application of field theory to the
physics of spin systems on a discrete lattice, and thus all necessary
regularizations are provided by the lattice in a natural way.)

Instead of working with the propagator (\ref{1to2}) in usual Minkovsky's
space-time, it is convenient to make the Wick rotation $t\to i\tau$
(essentially this procedure is an analytical continuation in $t$), passing to
the Euclidean space-time. Then one has the Euclidean propagator
\[
P_{AB}^{\rm eucl}=\langle
\varphi_{A}|e^{-\widehat{H}\tau_{0}/\hbar}|\varphi_{B}\rangle =\int {\cal
  D}\varphi\,\exp\big\{ -{\cal A}_{\rm eucl}/\hbar\big\} \,.    
\]
The main contribution to the path integral comes from the global minimum of
the Euclidean action functional ${\cal A}_{\rm eucl}$. This minimum
corresponds to a trivial solution $\varphi=\varphi_{0}=\mbox{const}$, where
$\varphi_{0}$ determines the minimal energy of the system.  However, if
several different values of $\varphi_{0}$ are possible, it is often important
to take into account the contribution from the {\em local} minima of the
Euclidean action as well. Such a local minimum can correspond, e.g., to a
trajectory $\varphi=\varphi_{\rm inst}(\tau)$ connecting two possible
$\varphi_{0}$ values; it is clear that the probability $P_{AB}$ will contain
the factor $\exp\{-{\cal A}_{\rm eucl}[\varphi_{\rm inst}]\}/\hbar$.  Such a
contribution can be calculated in a semiclassical approximation and describes
effects which cannot be accessed by means of the perturbation theory.

\begin{figure}
\mbox{\hspace{6mm}\psfig{figure=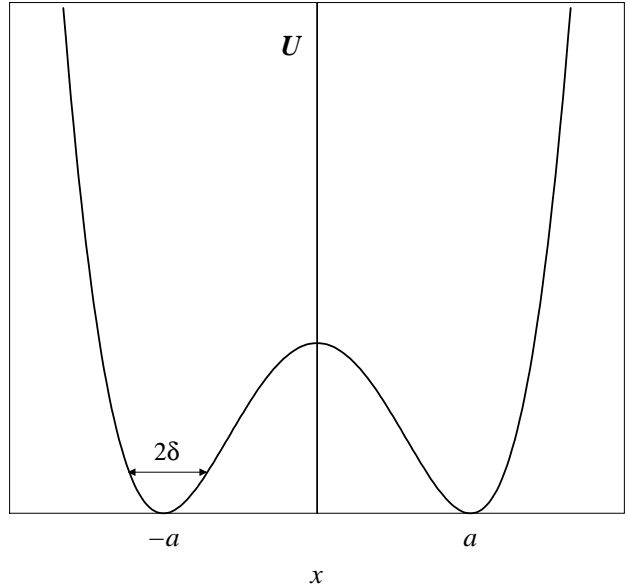,width=110mm,angle=-90.}}
%\vskip -9mm
\caption{A two-well potential $U(x)$ with equivalent wells at $x=\pm a$.
  Semiclassical treatment of tunneling
  is possible if the amplitude of zero-point oscillations $\delta\ll a$.}
    \label{fig:two-well}
\end{figure}

We will illustrate the above arguments on the example of a simple 
quantum-mechanical problem. Consider the motion of a particle of mass $m$ in a
symmetric two-well potential $U(x)$ of the type shown in Fig.\
\ref{fig:two-well},  with two equivalent minima at $x=\pm a$. Following
the popular choice \cite{inst}, we will assume this potential in the form
\begin{equation}
  \label{fi4}
  U(x)=\lambda (x^{2}-a^{2})^{2}\,,
\end{equation}
where the parameters $\lambda$ and $a$ determine the height and width of the
barrier between two wells. This model is described by the Lagrangian
\begin{equation}
  \label{Lfi4}
  L={m\over2}\left({dx\over dt}\right)^{2}-U(x)\,.
\end{equation}
After passing to the imaginary time, the Euclidean action is easily obtained
in the form
\begin{equation}
  \label{fi4A}
{\cal A}_{\rm eucl}=\int_{0}^{\tau_{0}} d\tau\, \Big\{ {1\over2}m
\left({dx\over d\tau}\right)^{2} + U(x) \Big\}\,.
\end{equation}
The classical (global) minimum of this functional is reached at $x=a$
or $x=-a$. Equations of motion for the action (\ref{fi4A})
\[
m{d^{2}x\over d\tau^{2}}={dU\over dx}
\]
correspond to the particle moving in the potential $-U(x)$, so that
$x=\pm a$ are {\em maxima} of this effective potential, and there exist
classical low-energy trajectories connecting them. Such trajectories represent
{\em local} minima of the Euclidean action functional and are called {\em
instantons}. They can be easily found in implicit form,
\begin{equation}
  \label{impl}
  \int dx\, \left({m\over 2U(x)}\right)^{1/2}=\tau-\tau_{0}\,,
\end{equation}
where $\tau_{0}$ is an arbitrary parameter determining the ``centre'' of
instanton solution. For many potentials the integration can be performed
explicitly, e.g., in case of (\ref{fi4}) one obtains
\begin{equation}
  \label{fi4inst}
  x=\pm a\tanh[\omega_{0}(\tau-\tau_{0})/2]\,,
\end{equation}
where $\omega_{0}=(8\lambda a^{3}/m)$ is the frequency of linear oscillations
around one of classical minima. Euclidean action for the instanton trajectory
can be written as
\begin{equation}
  \label{fi4A0}
  {\cal A}_{0}=\int_{-a}^{+a} \sqrt{2mU(x)}\, dx\,.
\end{equation}
For the model (\ref{fi4}) one has ${\cal A}_{0}=8a^{3}\sqrt{\lambda m}/3$.
Thus, instantons are very much like solitons with the difference that they are
localized in time.  Trajectories (\ref{fi4inst}) begin at $\tau\to-\infty$ in
one of the minima of $U(\varphi)$ and end at $\tau\to+\infty$ in the other
one; the contribution of those trajectories is responsible for the tunneling
splitting of the lowest energy level in the two-well potential. Indeed, the
tunneling level splitting is proportional to the matrix element $t_{12}$ of the
transition from one well to the other, and the probability amplitude of such a
transition is given by the path integral from $x=a$ to $x=-a$. It
is thus clear that the contribution of a single instanton to the transition
amplitude is proportional to $e^{-{\cal A}_{0}/\hbar}$.

The full calculation of this amplitude, however, is more complicated and
should take into account not only the instanton trajectories but all
trajectories close to them. Further, the full variety of multiinstanton paths
which bring the particle from one well to the other should be taken into
account. If the problem is semiclassical, i.e.\ ${\cal A}_{0}/\hbar$ is large
and the probability of tunneling is small, integration over ``close''
trajectories can be described as an effect of small fluctuations above the
instanton solution. Even this, usually elementary, problem of integrating over
small (linear) fluctuations is nontrivial in case of instantons, because some
of those fluctuations do not change the action. Particularly, from
(\ref{fi4inst}) it is easy to see that changing the position of instanton
centre $\tau_{0}$ has no effect on ${\cal A}_{0}$. Such ``zero modes'' {\em
always\/} arise in instanton problems and their contribution requires a
special analysis.  Detailed description of this technique would take us out of
the space limits, and we refer the interested reader to textbooks and review
articles (see, e.g., \cite{rajaraman,inst}).

We will attempt to get the correct result for the probability amplitude
$P_{AB}$ by means of the ``traditional'' quantum mechanics (without use of
path integrals and instantons). First, let us note that, due to the symmetry
of the potential $U(-x)=U(x)$, two lowest levels correspond to even and odd
eigenfunctions $\psi_{s}(x)$ and $\psi_{a}(x)$, with the energies $E_{s}$ and
$E_{a}$, respectively. Multiplying the Schr\"odinger equation for $\psi_{s}$
by $\psi_{a}$ and vice versa, then taking the difference of those two
equations and finally integrating over $x$ from $0$ to $\infty$, one obtains
the relation
\begin{equation}
  \label{rel1}
(E_{a}-E_{s})\left[\psi_{s}{d\psi_{a}\over dx}\right]_{x=0}=
\int_{0}^{\infty} \psi_{a}\psi_{s}\, dx\,,
\end{equation}
which is {\em exact\/} and is nothing but a mere consequence of the symmetry
properties. 

It is natural to try to use a semiclassical approximation. The semiclassical
result is given, e.g., in a popular textbook by Landau and Lifshitz
(\cite{LL-QM}, see the problem 3 after \S 50). According to that result,
$E_{s}-E_{a}=(\hbar\omega_{0}/\pi)\exp\{-{\cal A}_{0}'/\hbar\}$, where
$\omega_{0}=[k/m]^{1/2}$, $k\equiv(d^{2}U/dx^{2})_{x=a}$ and ${\cal
A}_{0}'=\int_{-a'}^{+a'}\big[2m\big(U(x)-E\big)\big]^{1/2}$, here $a'$ is the
turnover point of the classical trajectory with energy $E$ (corresponding to a
non-split level) defined by the equation $U(a')=E$. However, this result is
not adequate for our problem, and it does not coincide with the result of
instanton calculation. The point is that, surprisingly, the problem of
tunneling from one classical ground state to another {\em is not
semiclassical:\/} semiclassical approximation cannot be directly applied to
the ground state wavefunction inside one well.

Therefore we will do as follows: let us represent the wavefunctions inside the
barrier region as symmetric and antisymmetric combinations of the WKB
exponents,
\begin{eqnarray}
  \label{WKB}
&&\psi_{s}={C_{s}\over\sqrt{|p|}}\cosh\left\{{1\over\hbar}\int_{0}^{x}|p| dx
    \right\}\,\nonumber\\ 
&&\psi_{a}={C_{a}\over\sqrt{|p|}}\sinh\left\{{1\over\hbar}\int_{0}^{x}|p| dx
    \right\}\,
\end{eqnarray}
where $|p|=\sqrt{2m[U(x)-E]}$. Those wavefunctions can be used inside the
entire barrier region, except narrow intervals $|x\pm a|<\delta$ near the well
minima, where  $\delta=(\hbar/m\omega_{0})^{1/2}$ is the amplitude of
zero-point fluctuations.

On the other hand, if the condition $a\gg\delta$ is satisfied, then for the
description of the wavefunction inside the well any reasonable potential
$U(x)$ can be replaced by the parabolic one, $U(x)\to (k/2)(x\pm a)^{2}$. Then
in ``non-semiclassical'' regions one may use well-known expression for the
ground state wavefunction of a harmonic oscillator,
\begin{equation}
  \label{harm}
  \psi\to (\pi\delta^{2})^{-1/4}\exp[(x\pm a)^{2}/2\delta^{2}]\,.
\end{equation}
Thus, in the regions $a^{2}\gg (x\pm a)^{2}\gg\delta^{2}$ both the expressions
(\ref{WKB}) and (\ref{harm}) are valid. Then, normalization factors $C_{s,a}$
can be determined from the condition of matching (\ref{WKB}) and (\ref{harm})
in the two above-mentioned regions, and after that the integration in
(\ref{rel1}) can be performed explicitly. After some amount of algebra the
tunneling level splitting can be represented in the form
\begin{equation}
  \label{tunnel-QM}
  E_{a}-E_{s}=4\hbar\omega
  \sqrt{2\over\pi}\exp\left\{\int_{0}^{a-\delta}dx\,\sqrt{U''(a)\over2U(x)}
  \right\} \exp\left\{ -{1\over\hbar}{\cal A}_{0}\right\}\,,
\end{equation}
where the quantity ${\cal A}_{0}=\int_{-a}^{+a}dx\,\sqrt{2mU(x)}$ coincides
with the Euclidean action for the instanton trajectory. 

One can see that the difference between the formula (\ref{tunnel-QM}) and the
usual semiclassical result consists in the pre-exponential factor containing
the integral of the type 
$\int dx U^{-1/2}(x)$. It is clear that the main contribution into this
prefactor comes from the region $x\sim a$, where the integral can be
approximated as $\int^{a-\delta}dx/|a-x|$, so that it diverges logarithmically
at $\delta\to 0$. Thus for any potential $U(x)$ the prefactor can be
represented in the form $\widetilde{C}(a/\delta)$ or, equivalently, $(C{\cal
A}_{0}/\hbar)^{1/2}$. Here $C$ is a numerical constant of the order of unity,
it can be easily calculated for any given potential $U(x)$. So, finally we
arrive at the following universal formula:
\begin{equation}
  \label{tunnel-inst}
  E_{a}-E_{s}=4\hbar\omega_{0}\left({2C\over\pi}\right)^{1/2} \left({{\cal
        A}_{0}\over\hbar}\right)^{1/2} \exp\left\{-{{\cal
        A}_{0}\over\hbar}\right\}\, .
\end{equation}
For the model potentials $U=\lambda(x^{2}-a^{2})$ and $U=2U_{0}\sin^{2}x$ the
value of $C$ is equal to $\sqrt{3}$ and $\sqrt{2}$, respectively.

The formulas (\ref{tunnel-QM},\ref{tunnel-inst}) give the desired result for
any two-well potential with sufficiently large barrier. The main feature of
this result is the presence of an exponentially small factor. The small
parameter of the MQT problem is $\hbar/{\cal A}_{0}$, which can be represented
as a ratio of the zero-point fluctuations amplitude to the distance between
wells, $(\hbar/{\cal A}_{0})\sim (\delta/a)^{2}$. The expression
$e^{-a^{2}/\delta^{2}}$ is non-analytical in the small parameter, and thus the
MQT phenomenon cannot be obtained in any order of the perturbation theory.  We
wish to emphasize that the correct result is roughly $({\cal
A}_{0}/\hbar)^{1/2}$ times greater than that following from ``naive''
semiclassical formula. This large additional factor appears due to the
contribution from the regions close to the minima of the potential, where the
motion is not semiclassic. Let us try to understand this in the instanton
language.

As we mentioned before, the small exponential factor $\exp(-{\cal
  A}_{0}/\hbar)$  arises immediately in the instanton approach; the main
problem is to compute the pre-exponential factor, which is determined by the
integration over all small deviations from the instanton solution. Those
deviations are of two types:  real fluctuations of the instanton structure,
which increase the Euclidean action, and ``zero modes'' which correspond to
moving the instanton centre. It is rather clear that ``nonzero'' modes have a
characteristic energy of the order of $\hbar\omega_{0}$, and that the quantity
$\omega_{0}$ has nothing to do with the zero mode. Thus, it is obvious that
the factor $\hbar\omega_{0}$ arises from the integration over all ``nonzero''
modes, and the large factor $({\cal
  A}_{0}/\hbar)^{1/2}$ arises due to the zero (in our case -- translational)
mode. Such a ``separation'' naturally arises in rigorous calculations
\cite{inst,rajaraman}. 

It is remarkable that the above result can be generalized to the case of much
more complicated problems involving space-time instantons (which, as we will
see later, is important for the problem of MQT in topological
nanostructures). For any instanton {\em all\/} nonzero modes yield a factor
like $\hbar\omega_{0}$, and {\em each\/} of the zero modes yields the factor
$({\cal A}_{0}/\hbar)^{1/2}$ \cite{inst,rajaraman}, so that the final result
can be reconstructed practically without calculations (up to a numerical
factor of the order of unity). 

To illustrate one more feature typical for tunneling problems, let us consider
another model \cite{rajaraman}: a particle of mass $m$ which
can move along the circle of radius $R$, so that its
coordinate is determined by a single angular variable $\varphi$,
$0\leq\varphi\leq 2\pi$,  in the two-well potential
\begin{equation}
  \label{pocU}
  U(\varphi)=U_{0}(1-\cos 2\varphi)\,.
\end{equation}
The model is described by the following
Lagrangian:
\begin{equation}
  \label{pocL}
  L={1\over2}mR^{2} \left({d\varphi\over dt }\right)^{2} - U(\varphi)\,. 
\end{equation}
The classical Lagrangian can be modified by adding the arbitrary full
derivative term, e.g.,
\begin{equation}
  \label{pocTOP}
  L\mapsto L+\gamma {d\varphi\over dt}\,,
\end{equation}
which of course does not change the corresponding {\em classical} equations of
motion. However, adding the full derivative (\ref{pocTOP}) changes the
definition of the canonical momentum conjugate to $\varphi$, which, as one can
easily check, leads to a considerable change in the Hamiltonian of the
corresponding {\em quantum-mechanical} system after canonical quantization:
for nonzero $\gamma$ the correct Hamiltonian would be
\begin{equation}
  \label{pocH}
  \widehat{H}={1\over2mR^{2}}\Big\{i\hbar{d\over d\varphi} + \gamma \Big\}^{2}
  + U(\varphi)\,.
\end{equation}
Thus, there is no one-to-one correspondence between classical and
quantum-mechanical systems: several quantum systems can have the same
classical system as a classical limit.

For this model problem the instanton trajectories can be written down
explicitly:
\begin{eqnarray}
  \label{pocI}
&&  \cos\varphi=\sigma_{i}  \tanh[\omega(\tau-\tau_{i})]\,,\\
&& \omega=(4U_{0}/mR^{2})^{1/2}\,. \nonumber 
\end{eqnarray}
where $\tau_{i}$ is the arbitrary parameter determining the instanton position
in the imaginary time axis and $\sigma_{i}=\pm1$ is the topological charge
distinguishing instantons and
antiinstantons; the instanton action is finite and is
given by ${\cal A}_{0}=(8mR^{2}U_{0})^{1/2}$.

The importance of the full derivative term (\ref{pocTOP}) can be most easily
understood in terms of instantons. Indeed, let us consider  the tunneling
amplitude $P_{12}$ from the $\varphi=0$ well to $\varphi=\pi$ one: it is clear
that the contribution to this amplitude is made equally by instantons (with
$\varphi$ changing from $0$ to $\pi$) and antiinstantons (with $\varphi$
changing from $0$ to $-\pi$). However, the term (\ref{pocTOP}) becomes an {\em
imaginary\/} part of the Euclidean action and leads to the additional factor
$e^{i\pi\gamma/\hbar}$ associated with the instanton contribution and a
similar factor $e^{-i\pi\gamma/\hbar}$ for antiinstanton paths.  Thus, the
resulting transition amplitude for nonzero $\gamma$ is modified as follows:
\begin{equation}
  \label{ampltop}
P_{12}= [ P_{12}]_{\gamma=0}\cos(\pi\gamma/\hbar)\,,
\end{equation}
where $ [P_{12}]_{\gamma=0} \propto \omega \left({{\cal
A}_{0}/\hbar}\right)^{1/2}e^{-{{\cal A}_{0}/\hbar}}$, according to the general
result described above.  One can see that for half-integer $\gamma/\hbar$ the
interference of instanton and antiinstanton paths is {\em destructive,\/} so
that at $\gamma=\pm{\hbar\over2},\pm{3\hbar \over2},\ldots$ the tunneling
between two wells is {\em completely suppressed.\/} This effect is essentially
{\em topological\/} because the topological charge appears in the answer: the
contribution of configurations with different topological charge is
different. The same result can be obtained directly by solving the
Schr\"odinger equation with the Hamiltonian (\ref{pocH}): for half-integer
$\gamma/\hbar$ it can be mapped to the Mathieu equation with {\em
antiperiodic\/} boundary conditions, and the corresponding energy levels are
known to be doubly degenerate \cite{batemanIII}, which also means absence of
tunneling.
 
\section{Field-Theoretical Description of a Small Ferromagnetic Particle}
\label{sec:FM}

In this section we consider the basic technique of field-theoretical
description for spin systems on the simplest example, namely a nanoparticle of
a ferromagnetic material. Such an object may be viewed as a zero-dimensional
magnetic system, because at very low temperature all spins in the particle can
be considered as pointing in the same direction. 

It is worthwhile to consider first the dynamics of a single spin $S$. In order
to obtain the effective Lagrangian describing the spin dynamics, it is
convenient to use a coherent state path-integral approach (see, e.g., the
excellent textbook by Fradkin \cite{fradkin}).  Let us introduce a set of
generalized coherent states \cite{perelomov86}
\begin{equation}
|\vec{n}\rangle=\exp\{ i\theta(\vec{n}\times
\widehat{\vec{z}})\widehat{\vec{S}}\} |m=S\rangle
\label{one-site}
\end{equation}
parameterized by the unit vector $\vec{n}(\theta,\varphi)$.  Here
$\widehat{\vec{z}}$ is a unit vector pointing along the $z$ axis, and
$|m\rangle$ denotes a spin-$S$ state with $S^{z}=m$. They form a
non-orthogonal `overcomplete' basis so that the following property, usually
called a resolution of unity,
holds:
\begin{equation}
\int {\cal D}\vec{n}\,|\vec{n}\rangle \langle\vec{n}|=1\,,
\label{complete}
\end{equation}
another useful property is that quantum average of $\widehat{\vec{S}}$ on
those coherent states is the same as of classical vector of length $S$:
\[
\langle \vec{n}| \widehat{\vec{S}} |\vec{n}\rangle = S\vec{n}  \,.
\] 
In case of $S$=1/2 those coherent states have a very simple form and are
general single-spin wavefunctions:
\[
|\vec{n}\rangle=\cos(\theta/2)|\uparrow\rangle
+\sin(\theta/2)e^{i\varphi}|\downarrow\rangle\;.
\]

We again start from the formula for propagator (\ref{1to2}) which is
essentially a definition of the effective Lagrangian.  Slicing the time
interval $[0;t_{0}]$ into infinitely small pieces $\Delta t=t_{0}/N$, and
successively using the identity (\ref{complete}), one can rewrite this
propagator in $\vec{n}$-representation as 
\begin{eqnarray}
P_{AB}
&=&\lim_{N\to\infty}\int d\vec{n}_0 d\vec{n}_1
\cdots d\vec{n}_N \langle A|\vec{n}_{0}\rangle
\langle \vec{n}_N|B \rangle \nonumber\\
&\times& \prod_{k=0}^{N-1}
\langle\vec{n}_k| e^{-i\widehat{H}\Delta t/\hbar}
|\vec{n}_{k+1}\rangle\;.
\label{trotter}
\end{eqnarray}

Passing to the function $\vec{n}(t)$ of the continuum variable $t$, one ends
up with the coherent state path integral (\ref{1to2}) where the action ${\cal
A}$ is determined by the effective Lagrangian
\begin{equation}
L_{\rm eff}={1\over2}i\hbar \left\{ \langle\partial_t\vec{n}|
\vec{n}\rangle -
\langle\vec{n}| \partial_t
\vec{n}\rangle\right\} -
\langle\vec{n}|\widehat{H}|\vec{n}\rangle\;.
\label{efflagr}
\end{equation}

It can be shown that the dynamical part of this Lagrangian has the form
\begin{equation}
  \label{berry}
  \hbar S(1-\cos\theta){d\varphi\over dt}\,;
\end{equation}
for arbitrary $S$ this calculation requires some algebra, but for the simplest
case $S={1\over2}$ it is straightforward. The expression (\ref{berry}) is
nothing but the Berry phase \cite{berry84} for adiabatic motion of a single
spin. 

It should be remarked that the presence of the full derivative term $\hbar
S(d\varphi/dt)$ is rather nontrivial and allows one to capture subtle
differences between integer and half-integer spins, as we will see below. For
example, consider a single spin $S$ in some crystal-field potential, with the
effective Hamiltonian
\begin{equation}
  \label{singspinH}
  \widehat{H}=KS_{z}^{2} -K'S_{x}^{2}\,,
\end{equation}
where $K,K'>0$ and the easy-plane anisotropy $K$ is much stronger than the
in-plane anisotropy $K'$. The Lagrangian is
\begin{equation}
  \label{singspinL}
  L=\hbar S(1-\cos\theta){d\varphi\over dt} -KS^{2}\cos^{2}\theta
  -K'S^{2}\sin^{2}\theta\cos^{2}\varphi\,.   
\end{equation}
There are two equivalent classical minima of the potential at
$\theta={\pi\over2}$, $\varphi=0$ and $\theta={\pi\over2}$, $\varphi=\pi$.
Paths with $\theta\approx \pi/2$ make the main contribution into the
tunneling amplitude, so that we can approximately set
$\theta={\pi\over2}+\vartheta$, $\vartheta\ll 1$, and expand in $\vartheta$ up
to quadratic terms in the Lagrangian; in the term proportional to
$\vartheta^{2}$ the $K'$ contribution may be neglected as small comparing to
the contribution of $K$. After that, the ``slave'' variable $\vartheta$ can be
excluded from the Lagrangian (``integrated out'' of the path integral) because
the corresponding equation of motion $\delta L/\delta\vartheta=0$ allows to
express $\vartheta$ through $\varphi$ explicitly:
\begin{equation}
  \label{slave}
  \vartheta=-{\hbar \over2KS} {d\varphi\over dt}\,.
\end{equation}
Substituting this solution into the original Lagrangian (\ref{singspinL}), one
obtains the effective Lagrangian depending on $\varphi$ only:
\begin{equation}
  \label{singspinLeff}
  L_{\rm eff}= \hbar S{d\varphi\over dt} +
  {\hbar^{2}\over4K}\left({d\varphi\over
      dt}\right)^{2}+K'S^{2}\cos^{2}\varphi\,.  
\end{equation}
We see that we end up with the Lagrangian of a particle on a circle from the
previous section, with the topological term $\gamma=\hbar S$.  For each path
where $\varphi$ changes from $0$ to $\pi$ there is a corresponding
antiinstanton path with $\varphi$ changing from $0$ to $-\pi$, and those paths
contribute to the tunneling amplitude with phase factors $e^{i\pi S}$ and
$e^{-i\pi S}$. For half-integer $S$ those contributions precisely cancel each
other, making the tunneling impossible. This is exactly in line with the
well-known Kramers theorem, which states that in absence of external magnetic
field all energy levels of a system with half-integer total spin should be
twofold degenerate.  One can also straightforwardly check that for a single
spin in magnetic field, i.e., for $\widehat{H}=g\mu_{B}H\widehat{S}_{z}$, the
correct energy levels can be obtained only with the full derivative term taken
into account.

Now we are prepared enough, finally, to consider the problem of tunneling in
a small ferromagnetic particle consisting of $N$ spin-$S$ spins. If we assume
that ferromagnetic exchange interaction is so strong that we may consider all
spins as having the same direction, then we come to the ``giant spin'' model
where the entire particle is described as a quantum-mechanical
(``zero-dimensional'') system with only two degrees of freedom $\theta$ and
$\varphi$. In fact, we should postulate that in our path integral, when
integrating over the coherent state configurations $\prod_{i=1}^{N} \otimes
|\vec{n}_{i}\rangle$, the main contribution comes from the subspace with all
$N$ vectors $\vec{n}_{i}$ replaced by the same vector
$\vec{n}(\theta,\varphi)$, and we take into account only configurations from
this subspace. Assuming that the crystal-field anisotropy has the form
(\ref{singspinH}), we come to essentially the same effective Lagrangian
(\ref{singspinLeff}), and the only difference is that Eq.\
(\ref{singspinLeff}) should now be multiplied by the total number of spins
$N$. The tunnel splitting of the ground state level, according to Eq.\
(\ref{ampltop}), is given by
\begin{equation}
  \label{splitFM}
  \Delta E = C (NS^{3})^{1/2}(KK'^{3})^{1/4} |\cos(\pi NS)|
  \exp\left\{-NS(2K'/K)^{1/2}\right\}\,,
\end{equation}
where $C$ is a numerical constant of the order of $1$. A remarkable property
of the result (\ref{splitFM}) is that presence of a large number $N$ in the
exponent can be to some extent compensated by smallness of the ratio
$K'/K$. However, when the in-plane anisotropy $K'\to0$, the splitting vanishes
(this reflects the fact that in uniaxial case tunneling is impossible because
of the conservation of the corresponding projection of the total spin; the
same is true for $K\to0$).  Another remarkable feature is that for
half-integer $S$ the finite splitting can be observed only in particles with
even number of spins $N$; since in any statistical ensemble $N$ fluctuates a
bit, this roughly means that only one half of all particles gives nonzero
contribution.

Statistical fluctuations of $N$ have another, more painful consequence: since
$N$ stays in the exponent, even small fluctuations of the total number of
spins in the particle lead to large fluctuations of the splitting. Moreover,
since $N$ scales as the third power of the linear size $L$, small fluctuations
of $L$ will be considerably enhanced in $N$. This may be crucial if one tries
to detect the splitting by means of some resonance technique: the initially
weak signal would be even more weakened by the strong broadening of the
resonance peak. Actually, many factors can prevent one from observing the
tunneling resonance, e.g., relaxation, temperature effects, etc.  Here we will
not at all touch the problem of relaxation because of its complexity; instead
of that we refer the interested reader to the review
\cite{CaldeiraLeggett83}. Taking into account the finite temperature effects
is also nontrivial, particularly because it requires changing the procedure of
taking averages in the path integral: statistical averages should be taken
simultaneously with quantum-mechanical ones. Roughly (and without taking into
account the temperature dependence of relaxation mechanisms) the effects of
finite temperature can be estimated with the help of the concept of a
characteristic temperature $T_{c}$ below which the effects of quantum
tunneling prevail over thermal transitions. Rough estimate for $T_{c}$ is
obtained from the comparison of the relative strength of two exponential
factors: thermal exponent $e^{-\Delta U/T}$ and tunneling exponent $e^{-{\cal
A}_{0}/\hbar }$, where $\Delta U$ is the height of barrier separating two
equivalent states and ${\cal A}_{0}$ is the corresponding instanton action,
then $T_{c}= (\hbar \Delta U/{\cal A}_{0})$.  It is easy to see that for
the ferromagnetic particle problem considered above
\begin{equation}
  \label{Tfm}
T_{FM}=S(KK'/2)^{1/2}\,,  
\end{equation}
i.e. the temperature of crossover from classical to quantum transitions is
in this case rather small since it is determined by weak (relativistic)
anisotropy interaction constants; for typical anisotropy values $T_{FM}$ is
about $0.1$~K.

\section{Quantum Tunneling in a Small Antiferromagnetic
  Particle\protect\footnote{Subsection \protect\ref{sec:DM} was 
      written together with Vadim Kireev.}} 
\label{sec:AFM}

\subsection{Continuum field model of antiferromagnet}
\label{sec:AFMmodel}

The problem of continuum field description of antiferromagnet (AFM) is more
complicated but also much more interesting than a similar problem for
ferromagnet. Antiferromagnet contains at least two different ``sublattices''
whose magnetizations compensate each other in the equilibrium state. Thus,
when choosing the coherent state wavefunction in the form $|\Psi
\rangle=\prod_i|\vec{n}_i\rangle$ as described above, one cannot any more
consider $n_{i}$ as a ``smooth'' function of the lattice site $i$. Let us
adopt the simplest two-sublattice model which, despite the fact that it may be
inadequate for a specific material, still allows one to demonstrate the
essential physics of antiferromagnetism. We assume that there are two
equivalent sublattices with magnetizations $\vec{M}_{1}(\vec{r})$ and $\vec{M}_{2}(\vec{r})$,
$|\vec{M}_{1}|=|\vec{M}_{2}=M_{0}$.  Then, when passing to the continuum
limit, one has to introduce smooth fields
$\vec{m}=(\vec{M}_{1}+\vec{M}_{2})/2M_{0}$ and
$\vec{l}=(\vec{M}_{1}-\vec{M}_{2})/2M_{0}$ describing net magnetization and
sublattice magnetization, respectively. They satisfy the constraints
$\vec{m}\vec{l}=0$, $\vec{m}^2+\vec{l}^2=1$, and we further assume that
$|\vec{m}|\ll |\vec{l}|$. The energy of AFM $W=\langle \widehat{H}\rangle$
then can be expressed as a functional of $\vec{m}$ and $\vec{l}$:
\begin{equation}
  \label{W}
  W[\vec{m},\vec{l}]=M_{0}^{2} \int dV\, \left\{ {1\over2}\delta \vec{m}^{2}
    +{1\over2}\alpha(\nabla\vec{l})^{2}+w_{a}(\vec{l})  
  -{g\over M_{0}}(\vec{m}\cdot\vec{H})\right\} \,.
\end{equation}
Here the phenomenological constants $\delta$ and $\alpha$ describe homogeneous
and inhomogeneous exchange, respectively, $\vec{H}$ is the external magnetic
field, $g$ is the Lande factor, the function $w_{a}$ describes the energy of
magnetic anisotropy, and we use the notation
$(\nabla\vec{l})^{2}\equiv\sum_{i}(\partial\vec{l}/\partial x_{i})^{2}$. The
magnitude of sublattice magnetization $M_{0}=g\mu_{B}S/v_{0}$, where $\mu_{B}$
is the Bohr magneton, $S$ is the spin of a magnetic ion, and $v_{0}$ is the
volume of the magnetic elementary cell.

As we learned from the previous section, the correct Lagrangian, suitable for
the quantum-mechanical treatment, has the form
\begin{equation}
  \label{Lafm}
   L=\sum_{i} \hbar S \left\{ (1-\cos\theta_{1i}){d\varphi_{1i}\over dt}
   + (1-\cos\theta_{2i}){d\varphi_{2i}\over dt}\right\} -
   W[\vec{m},\vec{l}]\,,
\end{equation}
where the angular variables $(\theta_{1i},\varphi_{1i})$ and
$(\theta_{2i},\varphi_{2i})$ determine the unit vectors describing the
orientation of spins in first and second sublattice, respectively.  Note that
we have kept intact the summation sign in the dynamical part of (\ref{Lafm}):
the reason is that the explicit expression for the Berry phase in the
continuum limit strongly depends on the details of the magnetic elementary
cell structure (which dictates the correct definition of $\vec{m}$ and
$\vec{l}$ and the procedure of passing to the continuum limit).

Under the assumption that $|\vec{m}|\ll|\vec{l}|$, the magnetization $\vec{m}$
can be excluded from the Lagrangian (\ref{Lafm}), and one obtains the
effective Lagrangian depending only on $\vec{l}$; after that step $\vec{l}$
can be regarded as a unit vector, $\vec{l}^{2}=1$. 

For example, in the simplest case of an antiferromagnet with only two
(equivalent) atoms in elementary magnetic cell the dynamic part of the
Lagrangian (\ref{Lafm}) can be written as
\begin{equation}
  \label{Berry-ml}
\int dV 2\hbar S\vec{m}\cdot(\vec{l}\times \partial\vec{l}/\partial t)\,,
\end{equation}
and the density of the effective Lagrangian takes the form
\begin{equation}
  \label{LeffAFM}
  {\cal L}=M_{0}^{2} \left\{ {\alpha\over 2c^{2}}
 \left({\partial\vec{l}\over\partial t}\right)^{2}
 -{\alpha\over2}(\nabla\vec{l})^{2} -\widetilde{w}_{a}(\vec{l})\right\}
 +{4\over\gamma\delta}
 \vec{H}\cdot\left(\vec{l}\times{\partial\vec{l}\over\partial t} \right)\,,
\end{equation}
where $\widetilde{w}_{a}$ is the anisotropy energy renormalized by the
magnetic field,
\begin{equation}
  \label{wa}
  \widetilde{w}_{a}=w_{a}+{2\over\delta M_{0}^{2}}(\vec{l}\cdot\vec{H})^{2}\,,
\end{equation}
$\gamma=g\mu_{B}/\hbar$ is the gyromagnetic ratio, and $c={1\over2}\gamma
M_{0}(\alpha\delta)^{1/2}$ is the limiting velocity of spin waves. Using
general phenomenological arguments, one can show \cite{andrmar80} that in case
of arbitrary collinear antiferromagnet the Lagrangian should have the form
similar to (\ref{LeffAFM}).

Other, more complicated interactions can be present in Eq.\ (\ref{W}).  In
some AFM materials (which are, strictly speaking, weak ferromagnets) the
so-called Dzyaloshinskii-Moriya (DM) interaction is possible. It can be
described by including the term $D_{ik}m_{i}l_{k}$ under the integration sign
in into (\ref{W}), where $D_{ik}$ is some tensor (which is not necessarily
symmetric or antisymmetric). The origin of the DM interaction is rather
nontrivial, and there is a number of ``selection rules'' excluding the
possibility of its existence, particularly the DM interaction cannot exist (i)
if there is an inversion center interchanging sublattices; (ii) if there is a
translation which interchanges sublattices, i.e. if the magnetic elementary
cell is larger than the elementary cell of the original crystal lattice.  It
can be shown \cite{kik90} that presence of the DM interaction can be taken into
account by the substitution 
\begin{equation}
  \label{HDM}
\vec{H}\mapsto \vec{\widetilde{H}}=\vec{H}-{1\over2} M_{0}\vec{D}
\end{equation}
in the Lagrangian (\ref{LeffAFM}), where the components of vector $\vec{D}$
are defined as $D_{i}= D_{ik}l_{k}$.

If there exists a sublattice-interchanging inversion center, another invariant
may be present in (\ref{W}), namely
$\mu_{i}(\vec{m}\cdot\partial\vec{l}/\partial x_{i})$ (here $\mu_{i}$ are
certain exchange constants). It is very important for the physics of AFM in
one dimension, as we will see later.
 
\subsection{Spin tunneling in antiferromagnetic nanoparticle}

In case of a small particle one can consider $\vec{m}$ and $\vec{l}$ as being
uniform throughout the particle, i.e.\ as not having any space
dependence. Then, the Lagrangian (\ref{LeffAFM}) takes the form
\begin{eqnarray}
  \label{LpartAFM}
  L &=& {\hbar NS\over\gamma H_{e}} \left\{ \dot{\theta}^{2}+\sin^{2}\theta
    \dot{\varphi}^{2}  
  + 2\gamma \dot{\theta}\,(\widetilde{H}_{y}\cos\varphi
    -\widetilde{H}_{x}\sin\varphi) \right.\\
  &+&\left. 2\gamma\dot{\varphi}\,
    [\widetilde{H}_{z}\sin^{2}\theta
    -\sin\theta\cos\theta(\widetilde{H}_{y}\sin\varphi
    +\widetilde{H}_{x}\cos\varphi)] \right\}- M_{0}^{2}\widetilde{w}_{a}
  \,,\nonumber 
\end{eqnarray}
where $N$ is the total number of magnetic elementary cells in the particle,
$H_{e}=\delta M_{0}/2$ is the exchange field, the dot denotes differentiation
with respect to time, and we used angular variables for the vector $\vec{l}$,
\[
l_{z}=\cos\theta,\qquad l_{x}+il_{y}=\sin\theta\,e^{i\varphi}\,.
\]
There is another possible effect, typical only for antiferromagnetic
particles: due to the boundary (surface) effects, the number of spins in two
sublattices can differ from each other. In that case the Lagrangian
(\ref{LpartAFM}) will contain the additional term
\begin{equation}
  \label{noncomp}
  \hbar \nu  S (1-\cos\theta)\,\dot{\varphi}\,,
\end{equation}
which is essentially the Berry phase of $\nu$ non-compensated spins. Such a
sublattice decompensation in fact should be present in any ensemble of
nanoparticles, so that $\nu$ has certain statistical variation.

The full Lagrangian (\ref{LpartAFM}) is rather complicated, and for the sake
of clarity we will consider separately the effects of field and DM interaction.

\subsubsection{Tunneling in presence of external magnetic field}

Consider a small AFM particle with easy-axis anisotropy 
\[
w_{a} = {1\over2}\beta(l_{y}^{2}+l_{z}^{2}) 
\]
in external magnetic field $H$ perpendicular to the easy axis.
Then the Euclidean action takes the form
\begin{eqnarray}
  \label{AH}
  {\cal A}_{\rm eucl}&=&-{\hbar NS\over \gamma H_{e}} \int d\tau \Bigg\{
  \left({d\theta\over d\tau}\right)^{2}+\sin^{2}\theta\left({d\varphi \over
  d\tau}\right)^{2} +2i\gamma H\sin^{2}\theta{d\varphi\over d\tau}
  \nonumber\\ 
  &+&\omega_{0}^{2}\left[\sin^{2}\theta\sin^{2}\varphi
  +(1+\gamma^{2} H^{2}/\omega_{0}^{2})\cos^{2}\theta\right]\Bigg\} \\
&+&i\hbar\nu S\int d\tau (1-\cos\theta){d\varphi\over d\tau}  \,  \nonumber
\end{eqnarray}
where $\tau$ is the imaginary time, and $\omega_{0}={1\over2}\gamma
M_{0}(\delta\beta)^{1/2}$ is the characteristic magnon frequency
($\hbar\omega_{0}$ is the magnon gap).

There are two equivalent states $A$ and $B$ with opposite direction of
$\vec{l}$ along the easy axis $Ox$, and obviously the most preferable
instanton path is given by $\theta=\pi/2$, $\varphi=\varphi(\tau)$. The
instanton solution for $\varphi$ is the same as in case of particle on a
circle, and one-instanton action is
\begin{equation}
  \label{AinstH}
  {{\cal A}_{0}\over \hbar} = {NS\over\gamma H_{e}}\left( 4\omega_{0}\pm 2\pi i
    \gamma H \right) \pm i\pi\nu S \,,
\end{equation}
where $\pm$ signs correspond to instantons and antiinstantons.
Thus, the tunneling amplitude $P_{AB}$ is proportional to
\begin{equation}
  \label{PH}
\left({4NS\omega_{0}\over\gamma H_{e}}\right)^{1/2}
\exp\left\{-{4NS\omega_{0}\over\gamma H_{e}}\right\} \cos\left\{ \pi\nu S
+2\pi NS(H/H_{e})\right\}\,,
\end{equation}
and the corresponding magnitude of tunneling level splitting (proportional to
$|P_{AB}|$) oscillates with the period $\Delta H=(H_{e}/2NS)$ when changing
the external field. This period $\Delta H$ may be rather small, for typical
values of the exchange field $H_{e}\sim 10^{6}$~Oe and the number of spins in
the particle $N\sim 10^{3}\div10^{4}$ one obtains $\Delta H\sim
10^{2}\div10^{3}$~Oe. The effects of this type were studied in
\cite{DuanGarg94,GolyshevPopkov95}. 

The result (\ref{PH}) illustrates also another remarkable
feature: in any experiment probing the response of the ensemble of AFM
nanoparticles at each $H$ there must be only one possible value of splitting
(i.e., only one peak in the low-frequency response) when the spin of magnetic
ions $S$ is integer; but if $S$ is half-integer then, since in any ensemble
$\nu$ arbitrarily takes even and odd values, for approximately one
half of all particles the phase of cosine in (\ref{PH}) is shifted by $\pi/2$,
and there should be {\em two} peaks at each $H$.

It is worthwhile to note that the real part of the one-instanton action, which
enters the exponent in (\ref{PH}), is proportional to $(K/J)^{1/2}$ (where $J$
and $K$ are the exchange and anisotropy constants) while the
corresponding quantity for ferromagnet, according to (\ref{splitFM}), does not
contain the exchange constant and is determined by the rhombicity
$(K'/K)^{1/2}$. One may conclude that tunneling in AFM particles is 
more easy than in FM; indeed, the characteristic crossover temperature below
which quantum effects dominate over thermal ones, for antiferromagnets is
\begin{equation}
  \label{Tafm}
  T_{AFM}\propto S(KJ)^{1/2}\,,
\end{equation}
which is much greater than for ferromagnets [cf. Eq.(\ref{Tfm})]; typically
$T_{AFM}$ is about $1\div3$~K. 

\subsubsection{Tunneling in presence of the DM interaction}
\label{sec:DM}

Consider the same small AFM particle from the previous subsection, but imagine
that the DM interaction in its simplest form is present, with the energy
given by
\begin{equation}
  \label{wd}
  w_{d}=d(m_{y}l_{z}-m_{z}l_{y})\,.
\end{equation}
Then the DM interaction leads to the contribution into the Lagrangian
(\ref{LpartAFM}) of the form
\begin{equation}
  \label{Ld}
 \Delta L_{d}={\hbar NS\over\gamma H_{e}}\cdot 2 H_{D}{d\over
   dt}(\sin\theta\cos\varphi)\,, 
\end{equation}
where $H_{D}=dM_{0}$ is the so-called Dzyaloshinskii field. This term will
contribute to the imaginary part of the Euclidean action (\ref{AH}), and 
as a result the cosine in (\ref{PH}) will be modified as
\begin{equation}
  \label{Pd}
  \cos\left\{ \pi\nu S
+2\pi NS (H/H_{e}) + 4NS (H_{D}/H_{e})\right\}\,.
\end{equation}
Thus, presence of the DM interaction alone also leads to effective change of
the Berry phase and lifts the degeneracy for odd $\nu$ and half-integer $S$.

\section{Spin Tunneling in Topological Magnetic
  Nanostructures\protect\footnote{Subsection \protect\ref{sec:rings} was
      written together with Vadim Kireev.}}
\label{sec:top}

As we mentioned before, one of the most difficult experimental tasks when
trying to detect the resonance on tunnel-splitted levels in small particles is
to prepare the ensemble of particles with very sharp size distribution: even
small fluctuations of size lead to large fluctuations of the tunneling
probability since they contribute to the power of exponent.  Preparing such an
ensemble requires high technologies and involves considerable difficulties.
One may think about some other, ``natural'' type of magnetic nanostructures to
observe spin tunneling phenomena in. One nice solution, which have actually
been used in experiment, is to use biologically produced nanoparticles
\cite{Awschalom+92}. 

Another possible way, proposed in \cite{ivkol94,ivkol95tun,GalkinaIv95}, is to
use {\em topologically nontrivial magnetic structures:\/} kinks in quasi-1D
materials, vortices and disclinations in 2D, etc. Such objects have required
mesoscopic scale (e.g., the thickness of a domain wall is usually about 100
lattice constants) and, since their shape is determined by the material
constants, they are identical to a high extent (up to a possible inhomogeneity
of the sample).

Here we consider several possible scenarios of tunneling in
topological nanostructures and show that their use has a number of advantages.

\subsection{Tunneling in a Kink of 1D Antiferromagnet}

Consider a one-dimensional two-sublattice antiferromagnet with rhombic
anisotropy described by the Hamiltonian
\begin{equation}
  \label{Hafm}
  \widehat{H}=J\sum_{i} \vec{S}_{i}\vec{S}_{i+1} +\sum_{i}\big[
  K_{1}(S_{i}^{z})^{2} +K_{2}(S_{i}^{y})^{2}\big]\,,
\end{equation}
where $i$ labels sites of the spin chain with the lattice constant $a$,
$K_{1}>K_{2}>0$ are the anisotropy constants (so that $Oz$ is the difficult
axis and $Ox$ is the easy axis), and $J$ is the exchange constant.  For
passing to the continuum field description one may introduce vectors $\vec{m}$
and $\vec{l}$ as $\vec{m}_k= (\vec{n}_{2k+1}+\vec{n}_{2k})/2$ and $\vec{l}_k=
(\vec{n}_{2k+1}-\vec{n}_{2k})/2$, where $\vec{n}$ are the unit vectors
describing the direction of spins (the parameters of the corresponding
coherent states, see the discussion in Sect.\ \ref{sec:AFMmodel}) above.
These fields live on the lattice with the double spacing $2a$, and it is easy
to see that the energy functional $W=\langle\widehat{H}\rangle$ contains the
term $\vec{m}\cdot\partial_{x}\vec{l}$.  Using the equation $\delta
L/\delta\vec{m}=0$, one may express $\vec{m}$ through $\vec{l}$ and its
derivatives and exclude it from the Lagrangian.  The effective Lagrangian
takes the following form:
\begin{equation}
L_{\rm eff}= \int {dx\over 2a} \left\{
{\hbar^{2}\over4J}
(\partial_{t}\vec{l})^{2} -JS^{2}a^{2}(\partial_{x}\vec{l})^{2}
-K_{1}S^{2}l_{z}^{2} -K_{2}S^{2}l_{y}^{2}\right\} +L_{\rm top} \,,
\label{Lafm1d}
\end{equation}
which represents a (1+1)-dimensional nonlinear $\sigma$-model with the
so-called topological term
\begin{equation}
  \label{Ltop}
L_{\rm top}= {1\over2}\,\hbar S \int dx \,
\vec{l}\cdot(\partial_{x}\vec{l}\times\partial_t\vec{l}) \,.
\end{equation}
It is easy to trace the origin of this term: because of presence
of$\vec{m}\cdot\partial_{x}\vec{l}$ in the energy, the expression for
$\vec{m}$ contains $\partial_{x}\vec{l}$ which after the substitution into the
Berry phase (\ref{berry}) yields the topological term.  In agreement with
general phenomenological result (\ref{LeffAFM}), the Lagrangian (\ref{Lafm1d})
is Lorentz-invariant, with the limiting velocity $c=2JSa/\hbar$.

A stable kink solution corresponds to rotation of vector $\vec{l}$ in the easy
plane $(xy)$:
\begin{equation}
  \label{kink}
  l_{x}=\sigma'\tanh(x/\Delta),
  \quad l_{y}={\sigma\over\cosh(x/\Delta)},\quad
  l_{z}=0\,, 
\end{equation}
where $\Delta=a(J/K_{2})^{1/2}$ is the characteristic kink thickness, and the
quantities $\sigma$ and $\sigma'$ may take the values $\pm1$. The topological
charge of the kink $\sigma'$ is determined by the boundary conditions and
cannot change in any thermal or tunneling processes. The situation is
different with the quantity $\sigma$ which determines the sign of $\vec{l}$
projection onto the ``intermediate'' axis $Oy$. Two states with $\sigma=\pm1$
are energetically equivalent; change of $\sigma$ is not forbidden by
any conservation laws and describes the reorientation of the macroscopic
number of spins $N\sim\Delta/a\gg1$ ``inside'' a kink, typically $N\sim
70\div100$. 

\begin{figure}
\mbox{\hspace{6mm}\psfig{figure=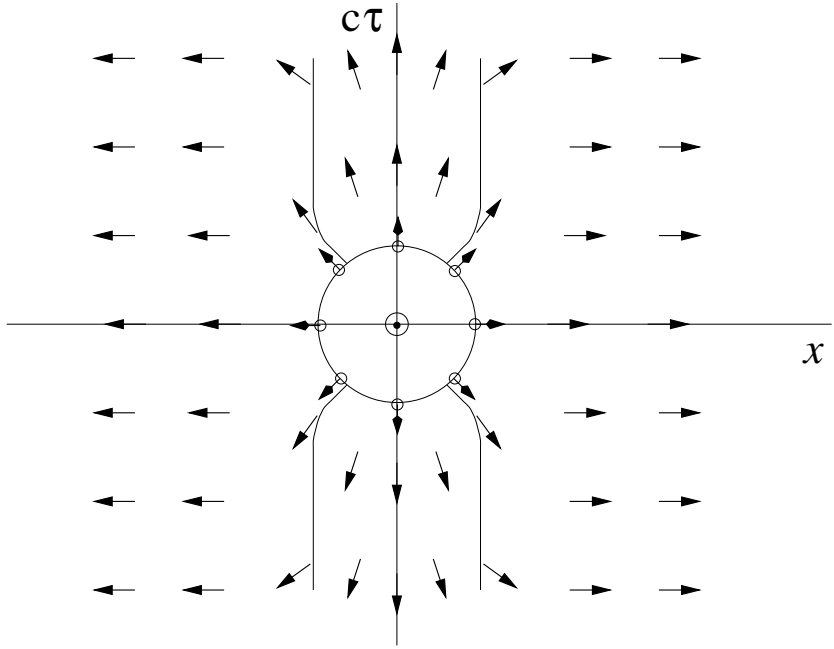,width=110mm,angle=-90.}}
  \caption{The structure of instanton solution for the problem of tunneling in
    a kink of a 1D antiferromagnet. Arrows and circles denote projections of
    vector $\vec{l}$ on the easy plane $(xy)$ and on the difficult axis $Oz$,
    respectively. Vector $\vec{l}$ forms the angle of about $45^{\circ}$ with
    the easy axis $Ox$ on thin solid curves, and with the difficult axis $Oz$
    on the circle (the circle radius is approximately $r_{0}$).}
  \label{fig:inst1}
\end{figure}

Again, tunneling between the kink states with $\sigma=\pm1$ can be studied
using the instanton formalism. In contrast to the case of a nanoparticle, here
the tunneling between two {\em inhomogeneous\/} states takes place, so that
nontrivial {\em space-time\/} instantons come into play. The instanton
solution $\vec{l}_{0}(x,\tau)$ is now {\em two-dimensional\/} and has the
following properties (see Fig.\ \ref{fig:inst1}):
\begin{eqnarray}
  \label{inst-prop}
  l_{x}\to\pm\sigma' && \mbox{at\ } x\to\pm\infty \nonumber\\
  l_{y}\to\mp\sigma && \mbox{at\ } x=0,\;\tau\to\pm\infty\\
  l_{z}=p=\pm1 && \mbox{at\ } x=0,\;\tau=0\,. \nonumber
\end{eqnarray}
Along any closed contour around the instanton center in the Euclidean plane
vector $\vec{l}$ rotates through the angle $2\pi \nu$ in the easy plane
$(xy)$, where $\nu=\sigma\sigma'=\pm1$. Thus, the instanton configuration has
the properties of a magnetic vortex and is characterized by two topological
charges \cite{affleck89rev,ivkol95rev}: vorticity $\nu$ and polarization
$p$. The instanton solution satisfies the equations
\begin{eqnarray}
  \label{inst-eq}
  &&\vec{\nabla}^2\theta +\sin\theta\cos\theta
  [(1+\rho\sin^2\varphi)/\Delta^2-(\vec{\nabla}\varphi)^2] =0,\nonumber\\
  &&\vec{\nabla}\cdot(\sin^2\theta\vec{\nabla}\varphi)
  -(\rho/\Delta^2)\sin^2\theta \sin\varphi\cos\varphi=0,
\end{eqnarray}
where we have introduced the angular variables $l_{y}+il_{z}=\sin\theta
e^{i\varphi}$, $l_{x}=\cos\theta$, $\rho=(K_{1}-K_{2})/K_{2}$ is the
rhombicity parameter, and $\vec{\nabla}=(\partial/\partial
x_{1},\partial/\partial x_{2})$ is the Euclidean gradient,
$(x_{1},x_{2})\equiv(x,c\tau)$.

Several important properties of the instanton can be obtained without using
the explicit form of the solution. First of all, note that this instanton has
{\em two\/} zero modes which correspond to shifting the position of its centre
along the direction of $\tau$ and $x$ axes, respectively. The physical meaning
of the first mode is the same as for 1D instanton, and the second mode
corresponds to moving the kink center in real space (the kink position in
infinite 1D magnet is not fixed in our continuum model); however, if the kink
center is fixed due to some effects (e.g., because of pinning on the lattice,
or by boundary conditions), so that the eigenfrequency of its oscillations is
comparable with the characteristic magnon frequency, then only one
zero-frequency mode is present.

The Euclidean action ${\cal A}_{\rm eucl}$ can be represented in the form
\begin{eqnarray}
  \label{Ainst}
&&  {\cal A}_{\rm eucl}={1\over2}S\hbar F +i2\pi S\hbar Q,\quad\mbox{where}
  \nonumber\\ 
&& F={1\over2}\int d^{2}x\,\big[(\vec{\nabla}\theta)^{2}
+\sin^{2}\theta(\vec{\nabla}\varphi)^{2}
+{1\over\Delta^{2}}\cos^{2}\theta\,(1+\rho\sin^{2}\varphi) \big] \nonumber\\
&& Q={1\over4\pi}\int d^{2}x\, \varepsilon_{\alpha\beta}\sin\theta
 \partial_{\alpha}\theta \partial_{\beta}\varphi\,.
\end{eqnarray}
Imaginary part of the Euclidean action is in this case completely determined
by the topological term $L_{\rm top}$. The word ``topological'' becomes now
clear, because $Q$ is the homotopical index of mapping of the $(x_{1},x_{2})$
plane onto the sphere $\vec{l}^{2}=1$ (the Pontryagin index, or the winding
number). For uniform boundary conditions at infinity in the $(x_{1},x_{2})$
plane $Q$ can take only integer values, but in our case
$Q=-p\nu/2=\pm{1\over2}$ is half-integer, which is typical for vortices (see,
e.g., \cite{affleck89rev,ivkol95rev}). For a kink with given $\sigma'$ there
are two instanton solutions with the same vorticity $\nu$ and different
polarizations $p$. Thus, the tunneling amplitude is proportional to $\cos(\pi
S)$ and vanishes when the spin $S$ of magnetic ions is half-integer. However,
the degeneracy can be lifted in presence of external magnetic field or the DM
interaction, as we will see below.

We are not able to construct the exact solution of Eqs.\ (\ref{inst-eq}), but
the estimate of the tunneling amplitude in various limiting cases can be
obtained from approximate arguments. For $\rho\ll 1$ the characteristic space
scale of $\varphi$ variation $\Delta/\sqrt{\rho}$ is much greater than the
kink thickness $\Delta$, and the problem can be mapped to one with a finite
number of degrees of freedom (one may introduce the variable $\phi$ having the
meaning of the angle of deviation out of the easy plane ``inside a kink'', so
that the instanton solution can be seeked in the form $\phi=\phi(\tau)$), then
it is easy to obtain \cite{ivkol95prl}
\begin{equation}
  \label{F-lowrho}
F\simeq 4\rho^{1/2}\quad \mbox{at\ } \rho\ll1\, .   
\end{equation}

In the opposite limiting case $\rho\gg 1$ one again has two different length
scales: the kink thickness  $\Delta$ and the ``core'' radius 
$r_{0}=\Delta(K_{2}/K_{1})^{1/2}$, $r_{0}\ll\Delta$. For $r\ll\Delta$ all
interactions except the exchange one can be neglected, and one may use the
``isotropic'' vortex solution
\begin{eqnarray}
  \label{iso}
&&  \theta=\theta_{0}(r),\quad \varphi=\nu\chi,\quad\nu=\pm1\,,\nonumber\\
&&  {d^{2}\theta_{0}\over dr^{2}} +\left({1\over\Delta^{2}} 
  -{\nu^{2}\over r^{2}}\right)
  +\sin\theta_{0}\cos\theta_{0} =0\,,
\end{eqnarray}
where $r=(x_{1}^{2}+x_{2}^{2})^{1/2}$, $\chi=\arctan(x_{2}/x_{1})$ are polar
coordinates in the $(x_{1},x_{2})$ plane. For $r\gg r_{0}$, i.e., far outside
the core, one can approximately assume that
\begin{equation}
  \label{far}
  \theta={\pi\over2},\quad
  \vec{\nabla}^{2}\varphi={\rho\over2\Delta^{2}}\sin2\varphi\,.
\end{equation}
Within a wide range of $r$ (for $r_{0}\ll r\ll \Delta$) the solutions
(\ref{iso}) and (\ref{far}) can be regarded as coinciding, and the integrand
in $F$ is proportional to $1/r^{2}$. 

Then, one may divide the integration domain into two parts: $r<R$ and $r>R$,
where $R$ is arbitrary in between $r_{0}$ and $\Delta$. For $r<R$ the solution
(\ref{iso}) may be used, yielding
$F_{r<R}=\pi\ln(\zeta R/r_{0})$ with $\zeta\simeq4.2$ \cite{KosVorMan83}. For
$r>R$, one can use a simple trial function approximately satisfying
(\ref{far}), e.g.,
\begin{equation}
  \label{far-app}
  \cos\varphi={x_{2}\over r}{1\over\cosh(x/\Delta)},\quad
  \sin\varphi={x_{1}\over r}{1\over\cosh(x/\Delta)}\,,
\end{equation}
which yields $F_{r>R}=\pi\ln(\zeta'\Delta/R)$ with $\zeta\simeq0.1$. Summing
up the two contributions, we obtain
\begin{equation}
  \label{F-largerho}
  F\simeq\pi\ln(0.42\Delta/r_{0})\quad\mbox{at\ } \rho\gg1\,. 
\end{equation}

The tunnel splitting of the ``ground state'' level of the kink 
\begin{equation}
  \label{splitK}
  \Gamma\propto 
\hbar\omega_{l}\,(FS/2)^{n/2}e^{-(FS/2)}\,|\Phi|\,, 
\end{equation}
where $\omega_{l}=2S(JK_{2}\rho)^{1/2}$ is the frequency of the out-of-plane
magnon localized at the kink, $\Phi$ is the factor determined by the imaginary
part of the Euclidean action [in the simplest model $\Phi=\cos(\pi S)$], and
$n$ is the number of zero modes which can be equal to $1$ or $2$ depending on
whether the kink position is fixed, see above.  It is easy to estimate the
crossover temperature for the problem of tunneling in a kink, comparing the
exponent in (\ref{splitK}) with $e^{-U_{0}/T}$, where $U_{0}\simeq
2S^{2}(\sqrt{JK_{1}}-\sqrt{JK_{2}})$ is the barrier height; for $\rho\gg1$
(i.e., $K_{1}\gg K_{2}$) and $n=1$ one obtains
\begin{equation}
  \label{Tkink}
  T_{k}\propto {S(JK_{1})^{1/2}\over\ln(K_{1}/K_{2})}\,,
\end{equation}
which is only logarithmically smaller than the corresponding temperature for a
particle (\ref{Tafm}). 

Let us discuss now the behavior of the imaginary part of the Euclidean action
in case of deviations from the simplest model (\ref{Hafm}) for which the
tunneling is prohibited for half-integer $S$. The most simple observation is
that in a spin chain with alternated exchange interaction, when along the
chain the strength of exchange constant alternates as
$J_{1}J_{2}J_{1}J_{2}\cdots$, the topological term (\ref{Ltop}) acquires
additional factor $J_{1}/J_{2}$ (see, e.g.,
\cite{affleck85,affleck-book}), which leads to $\Phi=\cos(\pi S
J_{1}/J_{2})$ and allows tunneling for half-integer $S$. Another way to lift
the degeneracy at half-integer $S$ is to ``switch on'' the DM interaction or
external magnetic field.

Consider the same model (\ref{Hafm}) with the addition of a magnetic field
$\vec{H}$ applied in the easy plane $(xy)$. Presence of the field leads to the
additional contribution to the imaginary part of ${\cal A}_{\rm eucl}$
\begin{eqnarray}
  \label{addH}
&&  {\cal A}_{\rm eucl}\mapsto {\cal A}_{\rm eucl}+i\hbar Q'\,,\nonumber\\
&& Q'={2S\over a}{H\over H_{e}} \int \vec{n}\cdot
\left(\vec{l}\times{\partial\vec{l}\over\partial x_{2}}\right) d^{2}x \,,
\end{eqnarray}
where $\vec{n}\equiv \vec{H}/H$. The mixed product in (\ref{addH}) can be
rewritten in angular variables as 
\[
-\sin\theta\cos\theta\, (n_{x}\cos\varphi+n_{y}\sin\varphi)\,
  {\partial\varphi\over \partial x_{2}} +(n_{y}\cos\varphi-n_{x}\sin\varphi)\,
  {\partial\theta\over \partial x_{2}}\,.
\]
One may note that $\sin\theta$ and $\theta$ significantly differ from zero
only in the vortex core, and thus the isotropic vortex solution (\ref{iso})
may be used for the calculation of $Q'$. After integration we obtain
\begin{eqnarray}
  \label{addQ}
&&  Q'=2S{H\over H_{e}}{\Delta\over a} p(A n_{x}+\nu B n_{y})\,,
  \nonumber\\ 
&& A=\int_{0}^{\infty}(dr/\Delta)\sin\theta_{0}\cos\theta_{0},\quad
B=\int_{0}^{\infty}(dr/\Delta)r(d\theta_{0}/dr)\,,
\end{eqnarray}
where $p$ and $\nu$, as earlier, denote the polarization and vorticity of the
instanton solution, and $A,B$ are numerical constants (recall that, according
to (\ref{iso}), the isotropic solution $\theta_{0}$ may depend only on
$r/\Delta$). After performing the summation in $p,\nu$, and with the account
taken of the contribution $Q$ coming from the topological term, the factor
$\Phi$ in (\ref{splitK}) will be modified as
\begin{equation}
  \label{PhiH}
  \Phi\mapsto\Phi_{H}=\cos\left(2ASn_{x}{H\over H_{e}}{\Delta\over a}\right)
\cos\left(\pi S + 2BSn_{y}{H\over H_{e}}{\Delta\over a}\right)\,,
\end{equation}
which means that for the given geometry only the field component perpendicular
to the easy axis lifts the degeneracy existing for half-integer $S$. Similarly
to the case of a small AFM particle, the tunneling amplitude is an oscillating
function of the external magnetic field $H$, but here the situation is more
complicated because the period of oscillations depends on the field
orientation. 

\subsection{Tunneling in Antiferromagnetic Rings with Odd Number of
  Spins}
\label{sec:rings}

\begin{figure}
\mbox{\hspace{6mm}\psfig{figure=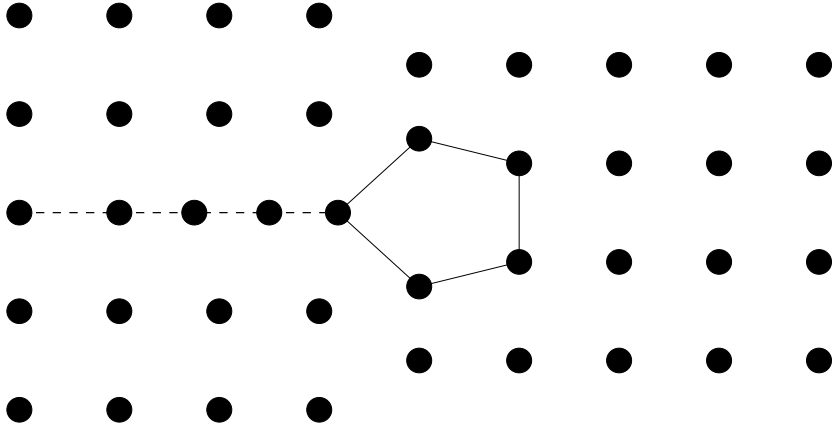,width=110mm,angle=-90.}}
  \caption{A ``ring'' around the core of dislocation in two-dimensional
    antiferromagnet. The dislocation is shown with a dashed line.} 
  \label{fig:ring}
\end{figure}

Another example of a magnetic nanostructure is a {\em ring\/} formed by
magnetic atoms; such rings may occur in a dislocation core of a 2D crystal as
shown in Fig.\ \ref{fig:ring}, and the characteristic feature of this object
is that the number of atoms in the ring is {\em odd}. Here we consider only
{\em antiferromagnetic\/} rings. In terms of the vector $\vec{l}$ such a ring
is a spin disclination. Let us assume that the magnetic anisotropy is of the
easy-plane type, and all spins lie in the $(xy)$ plane,
\[
\vec{S}_{i}=(-1)^{i}(\vec{e}_{x}\cos\varphi_{i}+\vec{e}_{y}\sin\varphi_{i})\,,
\]
where $\vec{e}_{x,y}$ are the unit vectors along $x,y$. Then there are two
energetically equivalent states of the ring, with $\varphi_{i}=\chi_{i}/2$ and
$\varphi_{i}=-\chi_{i}/2$, where $\chi_{i}$ is the azimuthal coordinate of the
$i$-th spin (let us assume that the ring is a circle of radius $R$). It is
possible to construct the instanton solution which links the two states; in
terms of $\vec{l}$ it can be written as
\begin{eqnarray*}
l_{x}=\cos{\chi\over2},\quad l_{y}=\sin{\chi\over2}\cos\psi,\quad
l_{z}=\sin{\chi\over2}\sin\psi\,, \\ 
\cos\psi=\pm\tanh(\omega_{0}\tau),\quad
\omega_{0}\simeq {1\over2}\gamma M_{0}(\beta\delta)^{1/2}\,.
\end{eqnarray*}
Calculation shows \cite{kireev} that the tunneling amplitude is proportional
to
\begin{equation}
  \label{splitR}
\cos(\pi S)\exp\{-\pi S R/\Delta \},\quad \Delta=(\alpha/\beta)^{1/2}\,,  
\end{equation}
i.e., the probability of tunneling is sufficiently large if the radius of the
ring is  smaller than the characteristic thickness of the domain wall $\Delta$
(usually $\Delta\sim100$\AA). Again, the tunneling is suppressed for
half-integer $S$, and this can be changed with the help of external magnetic
field. More detailed analysis \cite{kireev} shows that the field $\vec{H}$
should be applied in the easy plane in order to lift the degeneracy, then the
cosine in (\ref{splitR}) will change into
\[
\cos\left(\pi S +\pi^{2}S{H\over4H_{e}}{R\over a}\right)\,,
\]
where $a$ is of the order of the lattice constant. For weak fields the above
expression describes just the Zeeman splitting of the ground state level of a
ring (recall that due to the odd number of spins the ring always has an
uncompensated total spin if $S$ is half-integer).

\subsection{Tunneling in a Magnetic Vortex of 2D Antiferromagnet}

One more example of a magnetic topologically nontrivial structure is {\em
magnetic vortex\/} in quasi-2D easy-plane antiferromagnet. Consider the system
described by the Hamiltonian
\begin{equation}
  \label{H2dafm}
  \widehat{H}=J\sum_{\langle i,j\rangle} \vec{S}_{i}\cdot\vec{S}_{j}
  +K\sum_{i} \big(S^{z}_{i}\big)^{2}\,
\end{equation}
where $K>0$ is the anisotropy constant, and $Oz$ is the difficult axis. In
terms of the angular variables for the antiferromagnetism vector,
$l_{z}=\cos\theta$, $l_{x}+il_{y}=\sin\theta\,e^{i\varphi}$, a vortex
corresponds to the solution
\begin{eqnarray}
  \label{vort}
&& \theta=\theta_{0}^{\pm}(r),\quad \varphi=\nu\chi+\varphi_{0}\,,\\ 
&& \theta_{0}^{\pm}(\infty)=\pi/2,\quad \theta_{0}^{+}(0)=0,\quad
\theta_{0}^{-}(0)=\pi\,,
\end{eqnarray}
where $\theta_{0}$ satisfies the equation from the second line of Eq.\
(\ref{far}), $x+iy=re^{i\chi}$, and the solutions $\theta_{0}^{+}$ and
$\theta_{0}^{-}$ have the same vorticity $\nu$ but different polarizations
$p=\cos\theta(0)=\pm1$. The vortex states with $p=\pm1$ are energetically
equivalent, and the transition between them corresponds to reorientation of a
macroscopic number of spins $N\sim (\Delta/a)^{2}$, where
$\Delta=a(K/4J)^{1/2}$ is the characteristic radius of the vortex core and
$a$ is the lattice constant. It is worthwhile to remark that such a transition
would be forbidden in ferromagnet because of the conservation of the
$z$-projection of the total spin $S^{z}$.

\begin{figure}
\mbox{\hspace{6mm}\psfig{figure=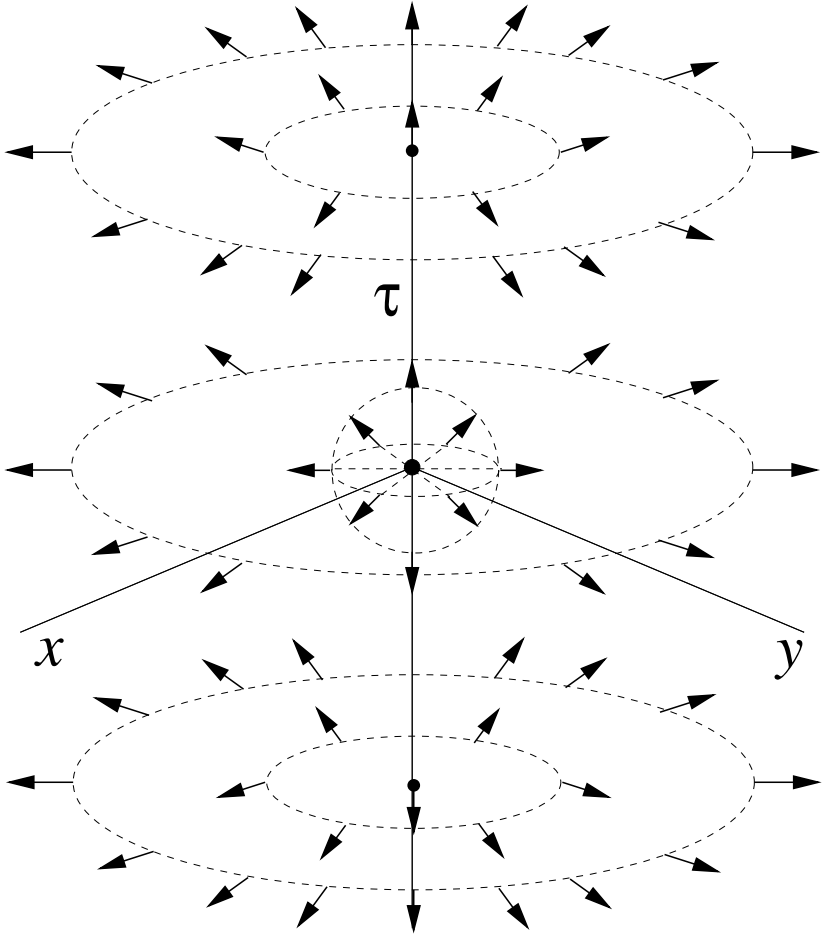,width=110mm,angle=-90.}}
  \caption{The structure of instanton solution for the problem of tunneling in
    a vortex. At $\tau\to\pm\infty$ one has $p=-1$ and $p=+1$ vortices,
    respectively. The sphere near the origin corresponds to the region where
    a hedgehog-type solution is adequate. }
  \label{fig:inst2}
\end{figure}

The instanton solution $\vec{l}(x,y,\tau)$ linking two vortex configurations
$\theta_{0}^{\pm}$ with $\nu=1$ when the imaginary time $\tau$ changes from
$-\infty$ to $+\infty$ is schematically shown in Fig.\ \ref{fig:inst2}. In the
3D Euclidean space $(x,y,\tau)$ it describes a topological configuration of
the hedgehog type and has a singularity at the origin. Such a singularity
means that in a small space region around the origin (roughly within the
distance of about $a$) one has to take into account the change of magnitude of
the sublattice magnetization: the length of the vector $\vec{l}$ has to change
so that $|\vec{l}(0,0,0)|=0$. In this case there are four zero modes, three of
them correspond to translations along $x$, $y$, $\tau$, and the fourth one
corresponds to changing the $\varphi_{0}$ angle. If the position and structure
of the vortex are fixed by some additional interactions, only one zero mode is
left. 

The Euclidean action derived from the Lagrangian of the $\sigma$-model has the
following form
\begin{equation}
  \label{Avor}
  {\cal A}_{E}= JS^{2}\int\! d\tau\!\int\! d^{2}x \left\{
{1\over c^{2}}\left({\partial\vec{l}\over\partial \tau}\right)^{2}
+(\vec{\nabla}\vec{l})^{2}- (\vec{\nabla}\vec{l}^{(0)})^{2}
+{1\over\Delta^{2}}\big[ l_{z}^{2}-(l_{z}^{(0)})^{2}\big]\right\} \,,
\end{equation}
where $\vec{l}^{(0)}$ describes the vortex solution (\ref{vort}) and $c$
denotes the limiting velocity $c=2JSa/\hbar$. Away from the singularity (for
$\rho\gg a$, $\rho\equiv (c^{2}\tau^{2}+x^{2}+y^{2})^{{1/2}}$) the condition
$\vec{l}^{2}=1$ holds, and the equations for $\theta,\varphi$ become
\begin{eqnarray}
  \label{inst2-eq}
  &&\vec{\nabla}^2\theta +\sin\theta\cos\theta
  [1/\Delta^2-(\vec{\nabla}\varphi)^2] =0,\nonumber\\
  &&\vec{\nabla}\cdot(\sin^2\theta\vec{\nabla}\varphi)=0\,.
\end{eqnarray}
In the region $a\ll \rho\ll\Delta$ this system has an exact centrally
symmetric solution of the hedgehog type:
\begin{equation}
  \label{hedgehog}
  \cos\theta={c\tau\over\rho}, \quad \tan\varphi={y\over x}\,.
\end{equation}
It can be shown that the contribution of the singularity itself is small and
can be neglected. Dividing the integration domain into two regions $\rho<R$
and $\rho>R$, where $R\ll\Delta$, one can see that the contribution of the
region of small distances $\rho<R$ to the Euclidean action is given by
\begin{equation}
  \label{lowrho}
  {\cal A}_{E}[\rho<R]=4\pi(JS^{2}/c)R\,.
\end{equation}
For estimating the contribution of the ``large'' distance region we use a
variational procedure with the trial function of the form
\begin{equation}
  \label{trial}
  \theta(x,y,\tau)=\pi/2 +F(c\tau)[\pi/2-\theta_{0}^{(+)}(r)]\,,
\end{equation}
where $F(c\tau)$ is a ``smeared step function:''  $F\to\pm1$ as
$\tau\to\pm\infty$ and the derivative of $F$ is nonzero in the region of the
thickness $\Delta_{1}$ around $\tau=0$. A simple estimate shows that the
resulting contribution of the region $\rho>R$ is described by
\begin{equation}
  \label{farrho}
  {\cal A}_{E}[\rho>R]=(2\pi JS^{2}/c)\big[
  \xi_{1}\Delta_{1}\ln(\Delta/R)
  +\xi_{2}\Delta_{1}+\xi_{3}\Delta^{2}/\Delta_{1} \big]\,,
\end{equation}
where $\xi_{1,2,3}$ are numerical constants of the order of unity. Summing up
(\ref{lowrho}) and (\ref{farrho}) and minimizing
${\cal A}_{E}$ with respect to $\Delta_{1}$ and $R$, we find
$\Delta_{1}\sim R\sim\Delta$. Thus, the total one-instanton Euclidean action
may be estimated as
\begin{equation}
  \label{Ainst2}
 {\cal A}_{0}=2\pi\xi JS^{2}\Delta/c=\xi\pi\hbar S\Delta/a\,, 
\end{equation}
where $\xi\sim 1$. 

Demanding that the tunneling exponent is not too large, e.g., ${\cal
A}_{0}<20\div30$, we see that for $S=5/2$ this means $\Delta/a<3\div4$, which
is rather tight; the continuum field approach we used here formally requires
$\Delta\gg a$, but in practice it is still applicable for $\Delta/a\sim2\div3$
\cite{galkina+93}. The crossover temperature $T_{c}\sim S(JK)^{1/2}$ is not
small since it is proportional to $\sqrt{J}$.

\section{Summary, and What is left under the carpet.}
\label{sec:summary}

Let us mention briefly the problems which are closely related to the topic of
this paper but were left out of discussion, and also those problems which are
not clear at present, to our opinion. 

First of all we would like to remark that we did not touch at all {\em
microscopic\/} essentially quantum effects in magnets, e.g., predicted by
Haldane destruction of (quasi)long-range order in 1D antiferromagnets with
integer spin $S$ caused by quantum fluctuations. Effects of quantum
interference are also important for this phenomenon, and its existence is
determined by the presence of topological term in the Lagrangian of
antiferromagnet (see, e.g., the reviews \cite{affleck89rev,affleck-book}). For
small $S$ and weak anisotropy the ground state of 1D antiferromagnetic system
can differ drastically from its classical prototype; e.g., the ground state of
a $S=1$ AFM ring is not sensitive to whether the number of spins is odd or
even and is always unique, and the ground state of a $S={1\over2}$ AFM ring
with odd number of spins is {\em fourfold\/} degenerate \cite{kireev}.

We also did not consider the contribution of tunneling-generated internal
soliton modes to the thermodynamics and response functions of 1D
antiferromagnets, which can lead to interesting effects (see
\cite{ivkol95prl,ivkol95rev,ivkol96jetp,ivkol97}).

Another problem which was ignored in our consideration is the role
of relaxation and thermal fluctuations of different origin. Even at low
temperature the interaction of spins with other crystal subsystems (lattice,
nuclear spins, etc.) may be very important, see
\cite{GargKim91,Garg93,Prokof'evStamp94}. It is clear that stochastic
influence on the dynamics of magnetization from thermal fluctuations leads to
decoherence and suppresses coherent tunneling. Description of this fundamental
problem in any detail goes far beyond the scope of the present lecture, and we
refer the reader to the review by Caldeira and Leggett
\cite{CaldeiraLeggett83}.

One more problem which is unclear from our point of view is a justification of
considering all spins in a small particle as moving coherently (a ``giant
spin'' approximation usually used in treating the MQT problems and also
adopted in the present paper). In fact, the only justification of this
approximation is energetical: if the particle size is much smaller than the
characteristic domain wall thickness, any inhomogeneous perturbation costs
much energy. On the other hand, for the Hamiltonian (\ref{singspinH}) neither
$\widehat{{\mathbf S}^{2}}$ nor $\widehat{S}^{z}$ are good quantum numbers,
which means presence of magnons (deviations from collinear order) in the
ground state.

Our lecture was devoted first of all to the fundamental aspects of MQT
considered as a beautiful physical phenomenon which is rather difficult to
observe. But technological development can lead to the situation when this
phenomenon will become practically important. The present tendency of
increasing the density of recording in the development of information storage
devices means decrease of the elementary magnetic scale corresponding to one
bit of information, and one may expect that quantum effects will determine the
``natural limit'' of miniaturization in future.

\section*{Acknowledgements}

Sections \ref{sec:DM} and \ref{sec:rings} were written together with V.~Kireev
whom the authors wish to thank for cooperation. This work was partially
supported by the grant 2.4/27 ``Tunnel'' from the Ukrainian Ministry of
Science.

\end{document}